\documentclass{WileyMSP-template}
\pdfoutput=1
\usepackage{xcolor}
\usepackage[T1]{fontenc}
\usepackage{multirow}
\usepackage{multicol}
\usepackage{amsmath}

\begin{document}

\pagestyle{fancy}
\rhead{\includegraphics[width=2.5cm]{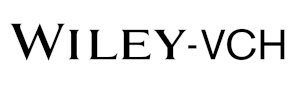}}

\title{Connecting Structure, Conformation and Energetics of Human\\ Telomere G-quadruplex Multimers}

\maketitle

% Author: Please give full first and last names for authors and include * after the name of all corresponding authors

\author{Benedetta Petra Rosi}
\author{Valeria Libera}
\author{Andrea Orecchini}
\author{Silvia Corezzi}
\author{Giorgio Schirò}
\author{Petra Pernot}
\author{Caterina Petrillo}
\author{Lucia Comez*}
\author{Cristiano De Michele*}
\author{Alessandro Paciaroni*}

% Affiliations: Please provide adacemic titles (Prof. or Dr.) for all authors where applicable, and include an institutional email address for all corresponding authors
\begin{affiliations}

Dr. B. P. Rosi, Dr. V. Libera, Prof. A. Orecchini, Prof. S. Corezzi, Prof. C. Petrillo, Prof. A. Paciaroni\\
Dipartimento di Fisica e Geologia\\
Università degli Studi di Perugia\\
Perugia 06123, Italy\\
E-mail: alessandro.paciaroni@unipg.it\\

\medskip
Prof. A. Orecchini, Dr. L. Comez\\
IOM-CNR c/o Dipartimento di Fisica e Geologia\\
Università degli Studi di Perugia\\
Perugia 06123, Italy\\
E-mail: comez@iom.cnr.it\\

\medskip
C. De Michele\\
Dipartimento di Fisica\\
Università degli Studi di Roma ``La Sapienza''\\
Rome 00185, Italy\\
E-mail: cristiano.demichele@uniroma1.it\\

\medskip
G. Schirò\\
Institut de Biologie Structurale\\
CNRS\\
Grenoble 38000, France\\

\medskip
P. Pernot\\
European Synchrotron Radiation Facility (ESRF)\\
Grenoble 38043, France\\
\end{affiliations}


\begin{abstract}

G-quadruplexes (G4s) are helical four-stranded structures forming from guanine-rich nucleic acid sequences, which are thought to play a role in cancer development and malignant transformation. Most current studies focus on G4 monomers, yet under suitable and biologically relevant conditions G4s undergo multimerization. Here, we address the structural, conformational and energetic features of G4 multimers formed in solutions by the human telomere sequence. A novel multi-technique approach is used combining Small Angle X-ray Scattering (SAXS) and circular dichroism experiments with coarse-grained simulations, to provide quantitative information about large-scale structural features and the stability of G4 multimers. The latter show a significant polydispersity with an exponential distribution of contour lengths, suggesting a step-growth polymerization. On increasing DNA concentration, the strength of G4 stacking interaction increases, as well as the number of the units in the aggregates, with dimers and trimers as the most probable forms. At the same time, a variation of G4 conformation is observed. Our findings provide a quantitative picture of human telomere G4 multimers, which must be accounted for to achieve a rational design of anticancer drugs targeting DNA structures.

\end{abstract}

\section{Introduction}
Deciphering the relationship between structure and function of biological molecules is a difficult but crucial task. This appears even more complex, if one considers that in the physiological context the role played by a biomolecule not only depends on the behavior in its monomeric state, but also on the interaction among different biomolecules and/or biomolecular units giving rise to higher-order structures. A paradigmatic case where the higher-order structure is crucial is represented by G-quadruplexes (G4s). These biomolecules are noncanonical nucleic acid structures formed by G-rich oligonucleotides which fold into four-stranded helical structures consisting of multiple stacked planar arrays of four guanine bases associated through cyclic Hoogsteen-like hydrogen bonds (G-tetrads) \cite{neidle2017}. G4s display three main topologies (parallel, antiparallel and hybrid) that differ in the relative orientation of the four guanine runs and in the arrangement of loop regions \cite{webba2007, neidle2003, karsisiotis2011}. They are highly polymorphic, as their structure depends both on the specific oligonucleotide sequence and on environmental factors, such as the type and concentration of cations, molecular crowding and/or dehydration conditions~\cite{huppert2010, phan2006, viglasky2011, chaires2010, smargiasso2008, 
heddi2011}. In the genomes of higher eukaryotes, sequences with the ability to form G4s are abundant \cite{huppert2005, todd2005, Haensel2017} and concentrated in the telomeric regions (up to 25 \% of all G4 DNA) \cite{biffi2013}. G4s have also been detected in cells \cite{lam2013, biffi2013}, where they are thought to regulate transcription, translation, DNA replication, RNA localization and other biological functions \cite{huppert2007, kendrick2010, rhodes2015}. Because of such biological importance, G4s have received a lot of attention as drug-design targets \cite{neidle2017,bianchi2018,comez2021}. In particular, since G4s have been demonstrated to block telomerase and HIV integrase, there is cause to believe that specific G4-stabilizing ligands could be used as anti-cancer or anti-viral drugs~\cite{neidle2010, perrone2014}.
In addition to this, G4s have been widely investigated as promising building blocks and functional elements in fields such as synthetic biology and nanotechnology \cite{yatsunyk2014, mergny2019}, mostly because of their high stability, structural versatility, and functional diversity.\\
The majority of studies on G4s have focused on their monomeric state, but there is evidence that G4s can take many different multimeric forms \cite{kolesnikova2019}. Aggregation has been shown to depend on the length of loops and likely to occur through the stacking of external G-tetrads of parallel folds, with dimers and trimers as the most probable aggregated forms \cite{smargiasso2008}.
It has been recently proposed that multimeric G4s may have an important biological role in the case of telomeric DNA, where the 3' single-stranded overhang has the capability to form higher-order structures containing several G4 units linked by TTA spacers \cite{monsen2021}. %Clustering several consecutive G4 units can take place thanks to interactions mediated by stacking of G-tetrads or overhanging nucleotides at the 5' or 3'-end of the sequence or by nucleotides in loops \cite{petraccone2013}. 
An effective design of anticancer drugs targeting telomere G4s \cite{zhao2020} clearly demands for a detailed knowledge of the spatial arrangement and stability of their multimeric structure. 
%However, even though these properties are quite well understood for monomeric G4s, their propensity to form aggregated forms is still poorly understood. 
Here we investigate the structural, conformational and energetic features of multimers formed by units of the human telomere sequence Tel22 [AG$_3$(T$_2$AG$_3$)$_3$], by exploiting a unique biophysical multi-technique approach that combines small angle X-ray scattering (SAXS), circular dichroism (CD) and coarse-grained simulations.
The comparison of experimental and numerical data allows a quantitative description of the multimerization of Tel22 G4s at increasing DNA concentrations and unveils how the aggregation process affects the structure, the conformation and stacking interactions of Tel22 G4s units.
%Such a multimerization is in turn associated with G4 structural and conformational changes which reflect on significant variations of the unit-to-unit stacking interaction, the volume of the G4 unit and the corresponding conformation, with a trend toward the parallel state.

\section{Results and Discussion}

\subsection{Small-angle X-ray Scattering}

\begin{figure} [h!]
  \includegraphics[width=3.35in]{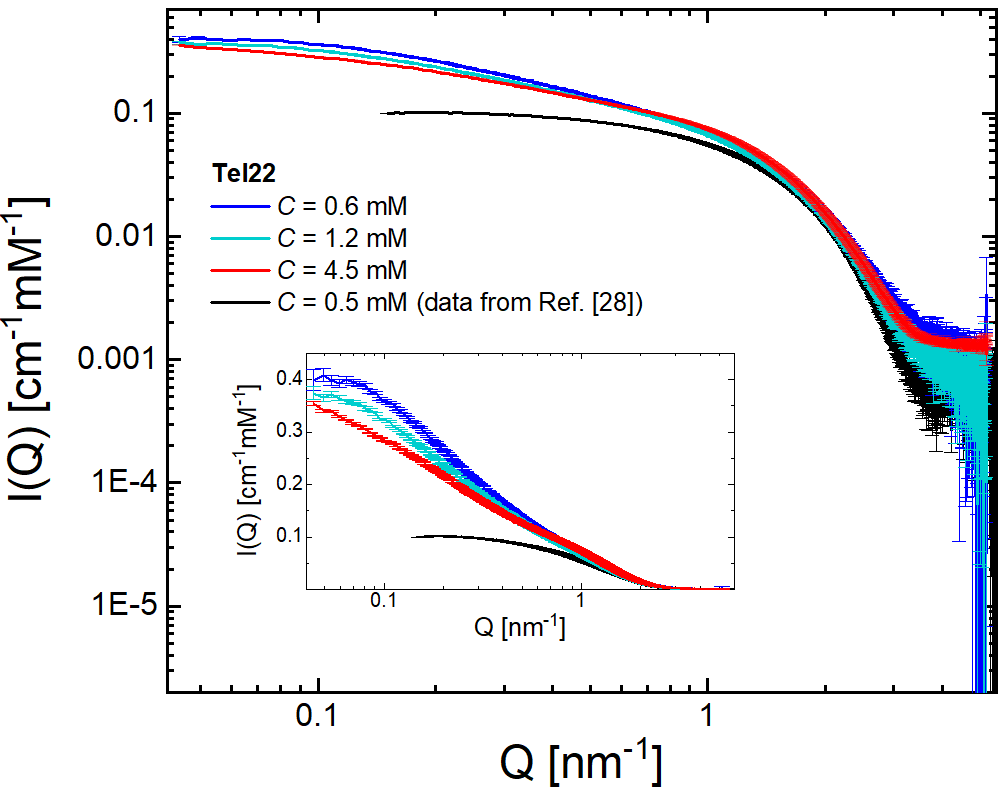}
  \caption{Logarithmic plot of the SAXS intensities of G4 solutions at different DNA concentrations. Data are reported in absolute scale and normalized to the molar concentration $C$ of DNA. For comparison, also the SAXS intensities of a Tel22 sample at $C$=0.5 mM are reported from Reference \cite{Libera2021}. In the inset, the same SAXS profiles are shown on a linear-log scale.}
  \label{fig:1}
\end{figure}

To experimentally investigate the structural properties of concentrated G4 samples, we performed SAXS measurements on G4 solutions at different DNA concentrations.
In the high-$Q$ region, $Q>1$ nm$^{-1}$, the signal accounts mainly for the structural features of the G4 unit (Tel22), i.e. shape and characteristic dimensions, through the so-called form factor function $P(Q)$. On the other hand, in the intermediate- and low-Q regions, the intensity $I(Q)$ reflects the higher-order structure arising from the reversible multimerization process of G4 units. In particular, the contribution of Tel22 multimers is quite visible, as a clear deviation below $Q<1$ nm$^{-1}$ in Figure \ref{fig:1}, where the scattering intensity of a solution of Tel22 monomers at DNA concentration $C=0.5$ mM (data taken from Reference \cite{Libera2021}) is compared with that of the present study. It is worth noting that the monomeric sample was prepared in such a way to rule out the presence of possible aggregates (see the Methods section).   
Actually, describing the complex mechanism of G4s self-assembly and deriving quantitative information on structural and conformational features of the G4 aggregates from SAXS measurements is a very difficult task, as we expect a significant degree of polydispersity in the length distribution of the aggregates \cite{Baldassarri2020}. Also in the high-Q region, where the SAXS signal is dominated by the form factor, the interpretation of the experimental data deserves great care. In a previous study~\cite{Libera2021}, we found the form factor of Tel22 monomers in solution to be well described by that of a hard cylinder (HC). In this case the topology of Tel22 consists of a mixture of antiparallel and hybrid conformers, however, an increase of concentration above $\approx 1$ mM drives a progressive shift toward the parallel (propeller) state, with concomitant decrease of the anti-parallel and hybrid forms \cite{Kejnovska2014,Vorlickova2012,Palacky2013,Renciuk2009}. This shift in the conformational populations entails a different description of the average G4 unit, as each conformer is characterized by a different shape and volume~\cite{An2014}. Since the only form factor known for the Tel22 sequence in solution is the one in the antiparallel conformation (143D structure in the PDB database \cite{wang1993}), we exploited numerical simulations in conjuction with experimental data to get new insights into the basic structural features of the different conformers.

\subsection{Simulation model}

\begin{figure} [h!]
  \includegraphics[width=3.35in]{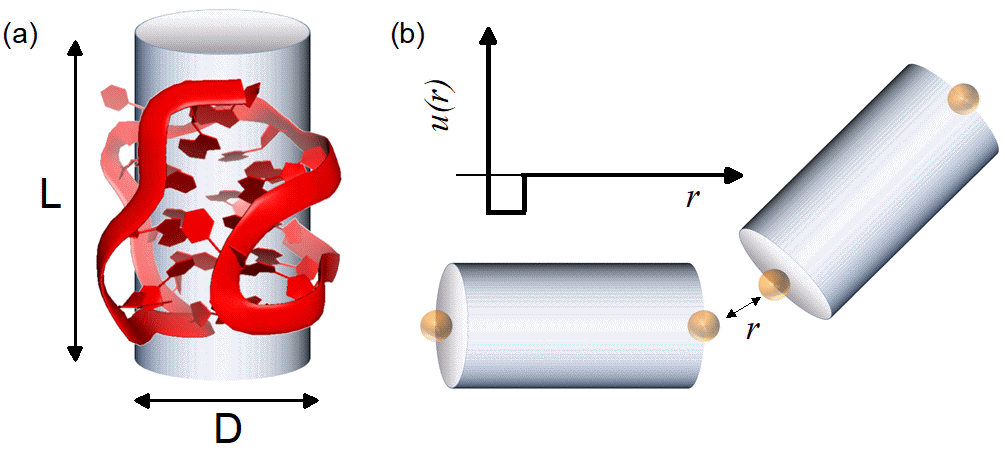}
  \caption{Simulation model for the G4 unit: (a) each G4 is modeled as a HC characterized by a diameter $D$ and a length $L$. (b) Each cylinder is decorated with two attractive sites at the basis. Sites belonging to different cylinders interact through the SW potential $u(r)$.}
  \label{fig:simul}
\end{figure}

To obtain more quantitative information on the G4 ensembles, we developed a novel approach 
comparing SAXS experimental intensities with those computationally estimated through Monte Carlo (MC) simulations, by modeling the G4 unit as a hard cylinder (HC) with two attractive sites at the basis, as schematized in Figure \ref{fig:simul} (for more details see the Methods section). Scattering intensities from simulations of HCs have been
obtained following a procedure similar to that discussed in  Reference~\cite{Pal2022}. Within this procedure each cylinder 
is replaced with a set of scattering points randomly placed inside its volume with a fixed number density, and by using these points
the structure factor is then calculated. This procedure ensures that the numerical scattering intensity also includes the form factor of HCs so that it can be directly compared with the experimental one. This simple yet effective coarse-grained model has proven to successfully describe reversible self-aggregation processes in other DNA-based systems \cite{Nguyen2015,Nguyen2014}, and it is especially suitable for describing the hydrophobic (stacking) forces acting between G4 units. 
An extensive campaign of simulations was carried out so as to reproduce SAXS results for all the concentrations studied experimentally. In particular, to find the best agreement between simulations and experiments we explored the phase space corresponding to different G4 shapes and different strength of stacking interaction between units. As to the former, the starting point was a HC with a diameter of $D_0=2.0$ nm and a length of $L_0=3.7$ nm that best reproduces the monomeric form of Tel22 sequences \cite{Libera2021}. To modulate the shape of the cylinder we introduced the parameter $K$, so that $D=D_0 \cdot K$ and $L=L_0/K$ ($K \geq 1$). Finally, the stacking interaction between G4s was varied through the effective temperature $T^*=k_BT/u_0$ (where $u_0$ is the binding energy of the HC attractive sites).
 
 \subsection{Comparison between SAXS data and simulations}

\begin{figure} [h!]
  \includegraphics[width=7.01in]{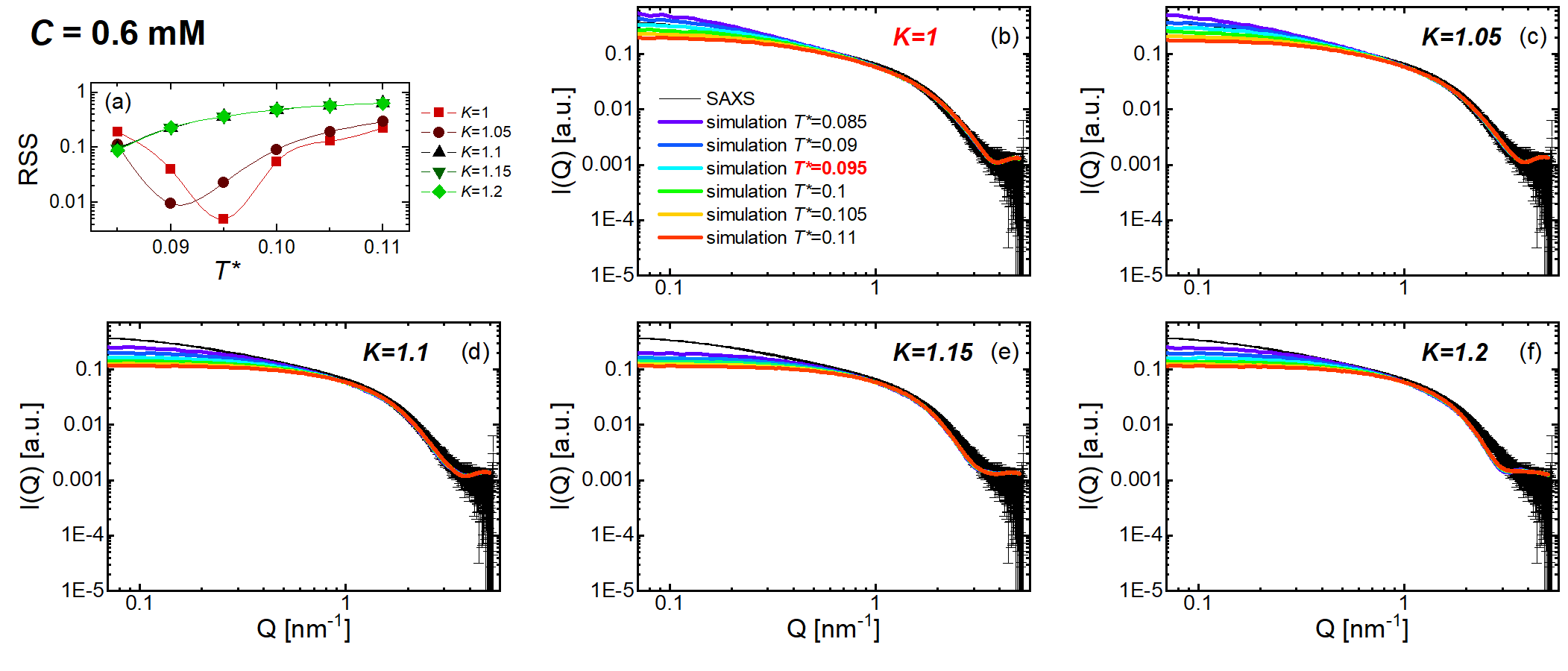}
  \caption{Scattering intensities for the G4 solution at $C$ = 0.6 mM, obtained from SAXS measurements and simulations (panels b -- f). The accordance between experimental and simulated data has been evaluated through the residual sum of squares ($RSS$), calculated at different values of $T^*$ and $K$ (panel a).}
  \label{fig:cfr}
\end{figure}

The experimental $I(Q)$ can be suitably reproduced by finely changing the $K$ and $T^*$ simulation parameters, as shown in Figure \ref{fig:cfr} for the selected concentration $C$=0.6 mM. In this case an excellent agreement between simulation and experiment is obtained for $K$=1 and $T^*$=0.095 (panel b). Similar plots for $C$=1.2 mM and $C$=4.5 mM are reported in Figure SI1 and SI2 of the Supporting Information, respectively.

\begin{figure} [h!]
  \includegraphics[width=3.35in]{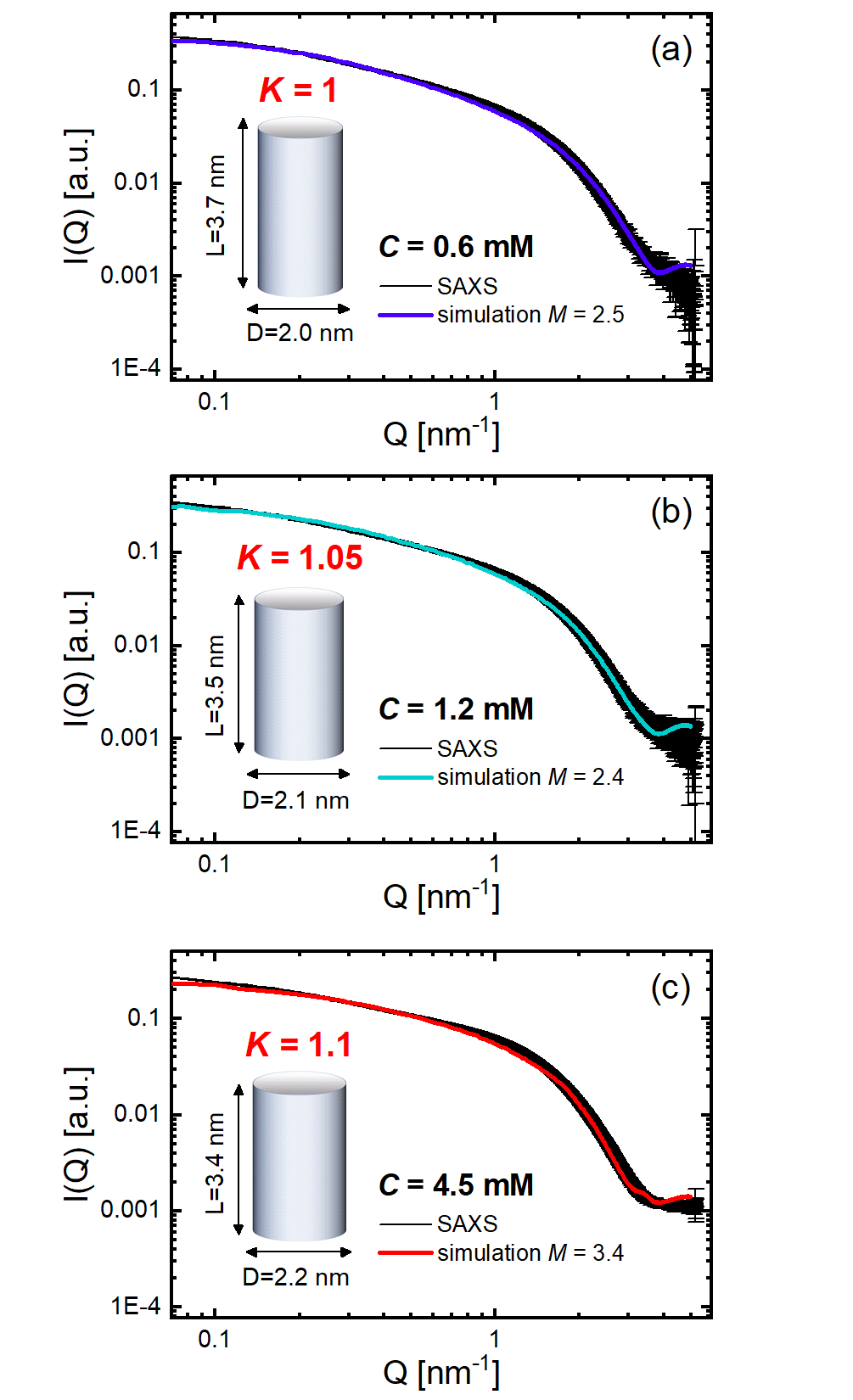}
  \caption{Panel a, b, and c of the figure show the best accordance between experimental and simulated scattering intensities, respectively for the G4 solution at $C$ = 0.6 mM, 1.2 mM, and 4.5 mM.}
  \label{fig:2}
\end{figure}

 Best matches (according to the residual sum of squares $RSS$) between experiments and simulations are reported in Figure \ref{fig:2}. Experimental and simulated curves remarkably superimpose onto each other over the whole explored $Q$-range. As expected, in the region of the highest $Q$ values (above $\approx 3 $ nm$^{-1}$) the simulated SAXS intensity is less effective in describing the experimental data, since it does not take into account the fine structural details of the actual G4 folds numerically modeled as HCs. On the other hand, simulations are able to reproduce strikingly well the experimental data in the low- and intermediate-$Q$ region, and can therefore effectively describe the aggregation processes.
 
  \begin{table}[h!]
 \centering
 \caption{List of parameters associated with the best representative state point ($K$,$T^*$) for each of the investigated concentration $C$.}
  \begin{tabular}{cccccccc}
    \hline
      $C$ [mM] & $T^*$ & $K$ & $M^a$ & \textit{\DH} &$G^0_{ST}$ [kcal mol$^{-1}$]$^b$ & $H^0_{ST}$ [kcal mol$^{-1}$]$^c$& $S^0_{ST}$ [cal mol$^{-1}$K$^{-1}$]$^d$\\
    \hline
    0.6  & 0.095  & 1 & 2.5 $\pm$ 0.3 & 1.60 $\pm$ 0.04 & -0.76 $\pm$ 0.17 & -6.12 $\pm$ 0.16 & -18.3 $\pm$ 0.8\\
    1.2  & 0.1  & 1.05 & 2.4 $\pm$ 0.2& 1.58 $\pm$ 0.04 & -0.70 $\pm$ 0.15 & -5.82 $\pm$ 0.15 & -17.5 $\pm$ 0.7\\
    4.5 & 0.09 & 1.1 & 3.4 $\pm$ 0.5& 1.70 $\pm$ 0.04 & -1.21 $\pm$ 0.19 & -6.47 $\pm$ 0.18 & -18.0 $\pm$ 0.9\\
    \hline
  \end{tabular}\\
\footnotesize$^a$Average chain length. Errors were estimated by considering the statistical error on $\epsilon$ and the error associated with the finite value $\Delta T^*$ by which temperatures were sampled in the simulations. 
$^b$Stacking free energy calculated for a standard concentration 1 M of G4s and $T=293$ K. $^c$Energy contribution to $G^0_{ST}$. $^d$ Entropic contribution to $G^0_{ST}$.
 \label{tab:1}
\end{table}
 
 Once validated, simulations can be exploited to obtain quantitative information on stacking and aggregation of G4 solutions as the concentration varies. In particular, the adopted numerical model is consistent with an exponential distribution $\nu(l)$ of G4 multimer chain lengths $l$: $\nu (l)= \rho M^{-(l+1)}(M-1)^{(l-1)}$, where $\rho = \sum_{l=1}^{\infty}l\nu(l)=N/V$ is the number of G4s per unit volume, and $M$ is the average number of stacked units. The value of $M$ can be directly obtained from the simulation, since it is related to the average potential energy per G4 monomer, $\epsilon$, through the relationship   $M=(1+\epsilon/u_0)^{-1}$~\cite{Sciortino2007}. The simulations thus provide an estimate of the average length of G4 multimers at each of the investigated concentrations, as reported in Table \ref{tab:1}. Quite interestingly, $M$ values are in the range 2.4 $\div$ 3.4, in agreement with the experimental results suggesting dimers and trimers as the main multimeric forms \cite{smargiasso2008}.\\
 It is worth noting that using an exponential distribution to describe the length of multimer chains is equivalent to assume that multimerization of G4s complies with a step-growth mechanism. This is analogous to the case of the self-assembly of DNA-encoded nanoparticles into chain-like superstructures \cite{Gu2019}, and at variance with the chain-growth mechanism which is less commonly applied to supramolecular biopolymers  \cite{Ogi2014,Zhang2019}. In addition, as Tel22 multimers follow an exponential distribution, we can easily estimate the spread of their molecular mass distribution as quantified by the ratio between the weight average molecular weight and the number average molecular weight, i.e. the so called dispersity {\textit {\DH}}$= 2-1/M$ \cite{Teraoka2002}. The parameter {\textit {\DH}}, which is reported in Table \ref{tab:1}, indicates a moderate degree of polidispersity as it is higher than the monodisperse limit {\textit {\DH}}$=1$ but smaller than the highest attainable value {\textit {\DH}}$=2$.\\
 
 The knowledge of the average chain length $M$ allows to calculate also the coaxial stacking free energy of the system $G_{ST}=-k_BT \ln [M(M-1)] $ \cite{DeMichele2012} (Table \ref{tab:1}). The standard free energy $G^0_{ST}$, calculated for a standard temperature of 293 K and concentration of 1 M, increases from about -0.7 kcal mol$^{-1}$ at the lowest concentrations to -1.2 kcal mol$^{-1}$ at $C=4.5$ mM. The associated contributions of the standard bonding energy $H^0_{ST}$ and the standard bonding entropy $S^0_{ST}$ are of the order of $\approx -6$ kcal mol$^{-1}$ and $\approx -18$ cal mol$^{-1}$K$^{-1}$, respectively. We note that our estimate for $G^0_{ST}$ is much lower than other values reported in literature for the stacking free energies of G4 structures. In a recent computational work, for instance, the dimerization free energy of parallel Tel22 G4s was estimated to be of the order of $-20$ kcal mol$^{-1}$ in the most stable 5'-5' configuration \cite{Kogut2019}. However, we point out that this disagreement can arise from the fact that free energy estimates based on all-atom molecular dynamics simulations may be quite inaccurate in calculating the entropic contribution to the stacking free energy, as it happens in the case of self-assembled short DNA-duplexes \cite{DeMichele2012,Maffeo2012,Maffeo2014,ChenPNAS2013}. For example, a system composed of $90$ ultra-short DNA duplexes of 5 base-pairs each has been studied through all-atom simulations in Reference~\cite{GlaserPRE17} and authors note that an overestimate of stacking
 attractions leads to unrealistically long aggregates. For comparison, computational studies in which only the enthalpic term is considered return G4s stacking energies between -34 and -8 kcal mol$^{-1}$ \cite{Spackova1999}, which are in line with our values. As current estimates of intermolecular G4 stacking energies mainly come from numerical studies, our findings call for future experimental work on this subject.\\

As for the geometry of the G4 units, the experimental curves are better reproduced using slightly larger values of $K$ on increasing concentration (as reported in Table \ref{tab:1}), corresponding to a variation of the simulated HCs towards a more disk-like shape. This trend suggests an increase of the fraction of G4 with parallel conformation, that is known to have a more oblate shape compared to the antiparallel and hybrid folds \cite{An2014}.  This is consistent with previous literature where the fraction of parallel G4s was proposed to progressively increase for DNA concentrations above $\approx 1$ mM \cite{Kejnovska2014,Vorlickova2012,Renciuk2009}. It is also worth of note that the parallel-stranded propeller topology might favour the stacking of G4 units \cite{Haider2008}, thus further 
favouring the formation of longer multimers on increasing concentration.
%explaining why we find longer oligomers at $C=4.5$ mM.\\ 

From the cylinder length $L$ and the average number of stacked units $M$ we can readily estimate the gyration radius of the G4 multimers, that turns out to be in the range 2.5 nm $\div$ 3.5 nm (see Supporting Information). To further support the picture provided by our method, 
in Figure SI3 gyration radii are shown to be in excellent agreement with those obtained by fitting the SAXS data with a phenomenological function proposed by Beaucage to describe systems with several characteristic lengths \cite{Beaucage1996}.

\subsection{Circular dichroism}

\begin{figure} [h!]
  \includegraphics[width=7.01in]{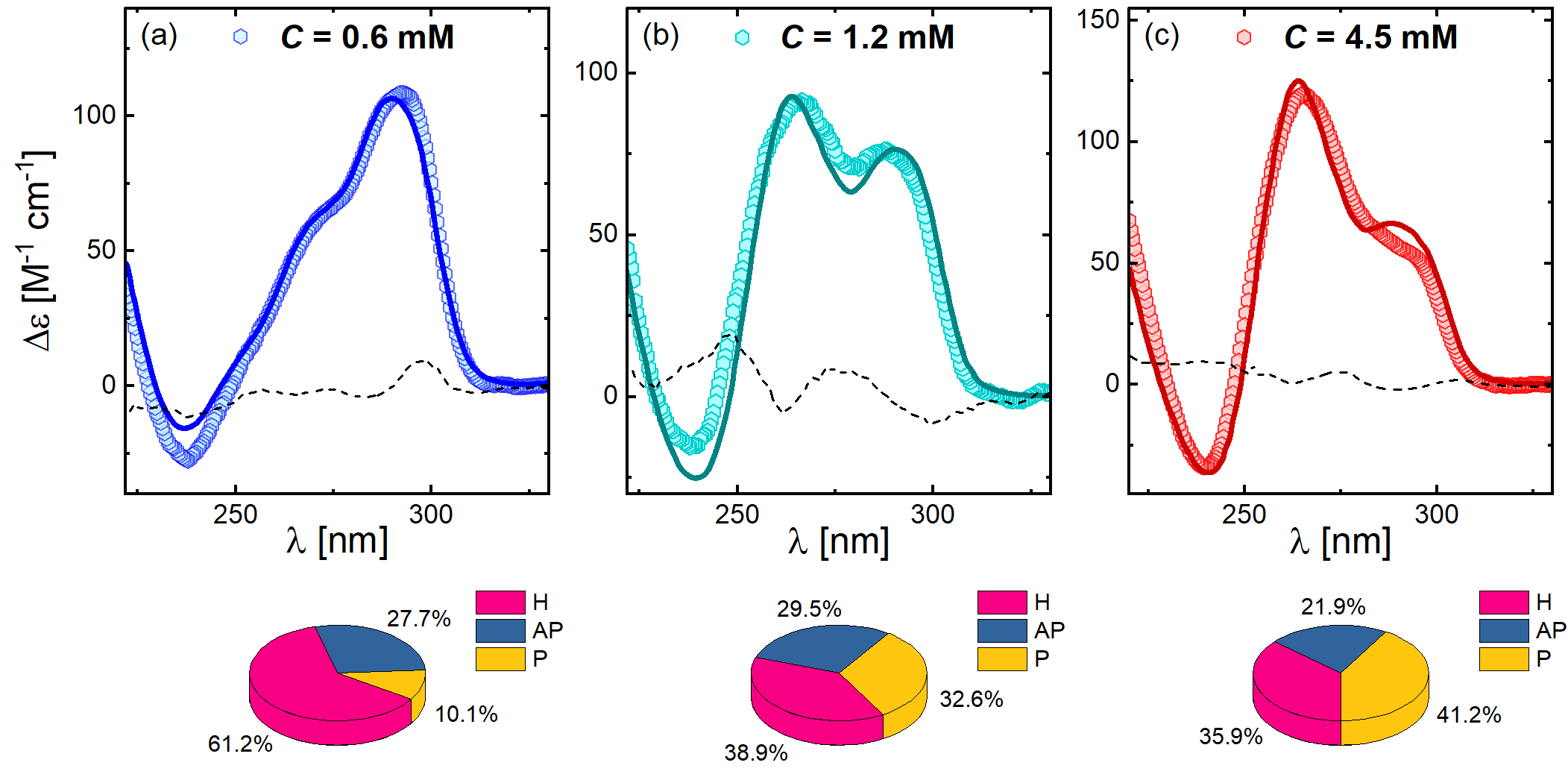}
  \caption{CD spectra of G4 solutions at different DNA concentrations (circles). In each panel, the fit resulting from the spectral deconvolution through the software by del Villar-Guerra \textit{et al.} is reported (solid lines), together with the corresponding residues (dashed lines). Pie charts in the bottom of each panel report the fraction of quadruplex units with antiparallel (AP), hybrid (H), and parallel (P) conformation, according to the spectral deconvolution.}
  \label{fig:5}
\end{figure}

To shed light on the possible change of the G4 secondary structure, we performed CD measurements as a function of the DNA concentration. As shown in Figure \ref{fig:5}, at $C=0.6$ mM (panel a) the spectrum displays the typical features of G4 hybrid topology, with a maximum at about 295 nm, a shoulder at 270 nm and a minimum at 240 nm. As $C$ increases, the maximum progressively downshifts and a new peak appears at 265 nm (panel b and c), suggesting an increase in the parallel fraction consistent with the findings reported above \cite{Renciuk2009}. A method based on the principal component analysis and singular value decomposition was used to quantify the fraction of the major G4 topologies \cite{VillarGuerra2018}. In Figure \ref{fig:5} the fit of experimental data obtained from the deconvolution into the three main folded topologies (parallel, antiparallel and hybrid), is reported, with the corresponding results represented in the pie charts. A population transfer is evident, as the hybrid component halves at the highest concentration, while the parallel component grows from 10\% to 40\%.

By combining the values of the G4 stacking energy and aspect ratio from SAXS and simulations with the fractions of topologies from CD, we can easily obtain specific information on the shape and energy parameters of each G4 conformation (see Table SI2 of the Supporting Information). As anticipated, a more elongated shape is associated with the antiparallel conformer, whereas a flattened cylinder represents the parallel one. As for the stacking interaction, it turns out that antiparallel G4s are much less prone to form multimers than either hybrid or parallel G4s (see Table SI2). This is consistent with previous results indicating that the parallel \cite{Kogut2019} and the hybrid  \cite{monsen2021} conformations are prevalent in human telomere multimers.\\

%Once the parameters associated with the individual G4 conformations are known, we can apply the same model to calculate the average chain length $M$ for aggregates of G4 with antiparallel, hybrid, and parallel conformation (as described in the Supporting Information). The analytical trends of $M$ as a function of the DNA concentration are reported in Figure SI4 of the Supporting Information. It should be pointed out that the predictive power of these trends is limited by the fact that, while for the mixtures the average value of the stacking energy is determined from the direct comparison between experimental data and simulations, for the single conformations such a value can be only indirectly estimated. Indeed, the exponential dependence of $M$ on stacking energy \cite{DeMichele2012} implies that even small deviations from the actual value of stacking energy lead to strong deviations in the value of $M$. By looking at Figure SI4, we can still qualitatively conclude that the tendency to form aggregates of the parallel and hybrid topologies is strongly favoured, compared to the antiparallel one.

\section{Conclusions}

This work provides an unprecedented quantitative description of the self-assembly process of Tel22 G4s in concentrated solutions.
Coarse-grained simulations were exploited to extensively explore the parameter space related to interaction stacking strength between G4s units, multimer length and the G4 unit shape. The coarse-grained simulation model was refined by successfully reproducing the experimental SAXS data, thus determining how the average length and the stability of G4 multimers changes as a function of DNA concentration. In the emerging picture, on increasing the self-crowding of G4s, the attractive interactions between different units become more and more important and at the same time the G4 conformation shifts toward the parallel state.\\
We show that our integrated approach allows a direct interpretation of low-resolution structural data on a challenging system consisting of polydisperse aggregates of polymorphic G4s, while also determining their thermodynamic features. In the light of this, we plan to adopt the same method in the future to provide a detailed description of G4s multimers even more closely related to biological issues, such as those forming from long telomeric sequences, in the presence of ligands of potential therapeutic interest. This knowledge will be crucial to design and optimize new drugs specifically targeting interunit junctions between adjacent G4s.

\section{Methods}
\subsection{Experimental Methods}

\subsubsection{Sample preparation}

Human telomere Tel22 AG$_3$(T$_2$AG$_3$)$_3$ ($M_w$: 6966.6 g mol$^{-1}$) was purchased from Eurogentec and used as received. Samples were prepared by dissolving the lyophilized powder in a 50 mM phosphate buffer at pH 7, 0.3 mM EDTA and 150 mM KCl. The high concentration of the starting solution ($\approx$ 13 mM) is appropriate to ensure the formation of aggregates, that remain stable even after dilution. The solution was heated up to 95°C, slowly cooled down to room temperature and left at room temperature overnight. After this procedure, the solution was centrifugated for 120 s at 15°C and 15000 rpm. From the centrifugated solution, samples at 3 different DNA concentrations were prepared, namely, $C$ = 0.6 mM, 1.2 mM, and 4.5 mM. The molarity of the solutions was determined from UV absorption measurements at 260 nm, using a molar extinction coefficient of 228 500 M$^{-1}$ cm$^{-1}$. Both experimental and computational investigations were performed at these concentrations. Before measurements, samples where further annealed and left at room temperature overnight.

It should be noted that the sample in monomeric state taken from Reference \cite{Libera2021} was prepared starting from a stock solution with a lower concentration of $C=1$ mM. Such a different procedure avoided the formation of aggregates.

\subsubsection{Circular Dichroism}

CD experiments were performed using a Jasco J810 spectropolarimeter in 0.01 to 0.1 mm path-length quartz cells. Spectra were collected in the range from 220 nm to 330 nm with a scan speed of 50 nm/min. All spectra were acquired at room temperature. CD data were expressed as the difference
in the molar absorption $\Delta \epsilon$ [M$^{-1}$ cm$^{-1}$] of the right- and left-handed circularly polarized light.

\subsubsection{Small-angle X-ray Scattering}

Small-angle X-ray Scattering (SAXS) experiments were performed at the BM29 beamline of the European Synchrotron Radiation Facility (ESRF) in Grenoble, France. The incident energy was 12.5 keV, corresponding to an incident wavelength of 0.99 Å$^{-1}$. The scattering vector range was between $Q$ = 0.0044 Å$^{-1}$ and 0.521 Å$^{-1}$. All patterns were collected at 20° C. Analogous patterns of the buffer were collected before and after every collection on the samples and used to subtract any contribution from the solvent and the sample environment. 

\subsection{Computational Methods}

\subsection{Simulation model}

The simulation model consisted of hard cylinders (HCs) characterized by a length $L$ and a diameter $D$, with two attractive sites at the bases. The attractive sites were located along the symmetry axis at a distance $L/2$ + $0.15D/2$ from the HC center of mass. Sites belonging to distinct particles interact via the square-well (SW) potential, i.e., $\beta u_{SW} = -\beta u_0$, if $r > \delta$, and $\beta u_{SW}$ = 0, if $r > \delta$, where $r$ is the distance between the interacting sites,  $\delta=0.25D$ is the interaction range (i.e., the diameter of the attractive sites), and $\beta u_0$ is the ratio between the
binding energy and the thermal energy $k_BT$, where $k_B$ i s the Boltzmann constant. The temperature was expressed as the adimensional parameter $T^*=k_BT/u_0$.

\subsubsection{Monte Carlo simulations}

Simulations were performed in the canonical ($NVT^*$) ensemble, leveraging a recently developed algorithm for checking the overlap between HCs~\cite{CDMEPJE18} which relies on a novel and very efficient algorithm for finding the roots of a quartic equation~\cite{CDMTOMS20}. For each value of concentration, we simulated a suitable number of particles $N$ in a cubic box with volume $V$ using standard periodic boundary conditions. Values of $N$ and $V$ where chosen so that the density $\rho=N/V=N_{av} C$, where $N_{av}$ is tha Avogadro's number, reproduced each of the experimentally-investigated values of DNA concentration $C$. The number of particles used in the simulations was $N=$ 5808, 6292 and 6776 for $C=$0.6 mM, 1.2 mM and 4.5 mM respectively. The aspect ratio of the HCs was varied from that associated with the starting values of $D_0=2$ nm and $L_0=3.7$ nm, by introducing the parameter $K$: $D=D_0 \cdot K$ and $L=L_0/K$. For each value of $C$, five different values of $K$ ($K$ =1, 1.05, 1.10, 1.15, and 1.2) were considered, thereby obtaining a total of 12 starting configurations. Each starting configuration was thermalised for at least $10^6$ MC steps at six different values of $T^*$, namely, $T^*$=0.085, 0.9, 0.095, 0.1, 0.105, and 0.11. The initial configuration used for the equilibration
was obtained by placing the hard cylinders on a ortorhombic lattice.
\medskip

\textbf{Supporting Information} \par %Please delete the Supporting Information statement if it is not applicable. Please supply Supporting Information in another file. Supporting information should not be provided in .tex format
Supporting Information is available from the Wiley Online Library or from the author.

% Acknowledgements
\medskip
\textbf{Acknowledgements} \par 
B.P.R. and C.P. acknowledges the support from MIUR-PRIN (Grant No. 2017RX9XXY). C.D.M. acknowledges the support from MIUR-PRIN (Grant No. 2017Z55KCW).

% References
\medskip

% Use the following code if you wish to generate your bibliography with BibTeX;
% replace the string "MSP-template" below with the name(s) of
% the BibTeX data base(s) you want to use.
% The resulting bibliography-output (the content of the .bbl file)
% must be pasted back into this file before submission.
% Please also include your BibTeX data base file(s) in your submission
% so that we can re-run BibTeX if necessary.
%
\bibliographystyle{MSP}
\bibliography{G4concentrated}

% Table of contents entry should be 50 - 60 words long
% Image should be 55 mm broad and 50 mm high or 110 mm broad and 20 mm high

%\medskip

%\begin{figure}
%\textbf{Table of Contents}\\
%\medskip
%  \includegraphics{toc-image.png}
%  \medskip
%  \caption*{ToC Entry}
%\end{figure}

\end{document}

\begin{figure}[t!]
\centering\includegraphics[width=4.5cm]{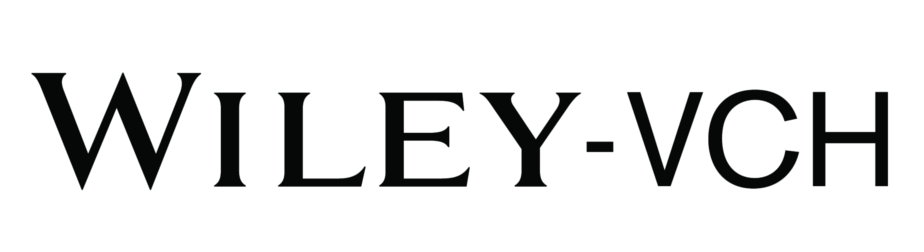}
  \end{figure}
  \vskip25pt
  \begin{center}
  {\Huge\sffamily\bfseries Supplementary Information for\par}
  \bigskip
  {\LARGE\sffamily\bfseries Connecting structure, conformation and energetics of human telomere G-quadruplex multimers\par}
  \bigskip
  {B. P. Rosi, V. Libera, A. Orecchini, S. Corezzi, G. Schirò, P. Pernot, C. Petrillo, L. Comez*, C. De Michele*, A. Paciaroni*\par\bigskip E-mail:  comez@iom.cnr.it; cristiano.demichele@uniroma1.it; alessandro.paciaroni@unipg.it\par}
  \end{center}

\subsection*{1. Comparison between experimental and simulated intensities for samples with $C=$ 1.2 mM and 4.5 mM}

\begin{figure} [h!]
\centering
  \includegraphics[width=5.7in]{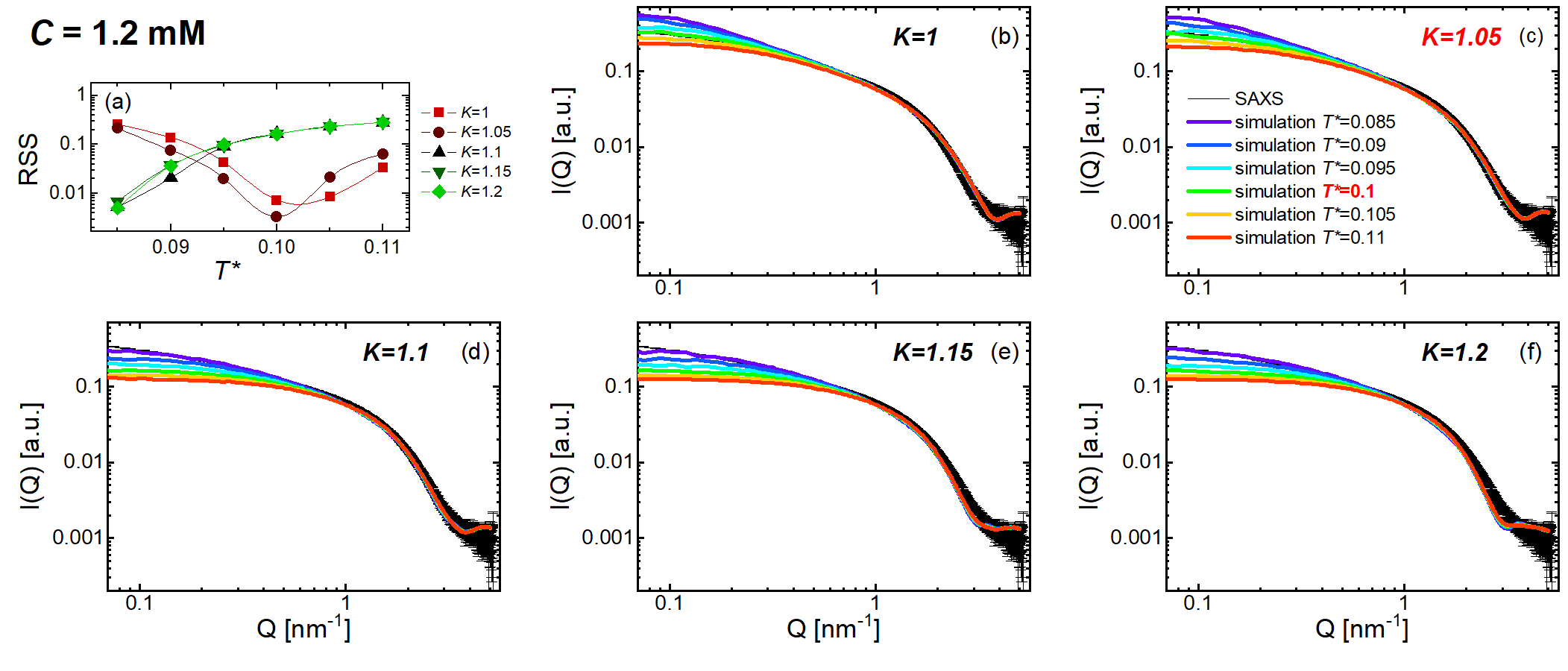}
  {\caption*{\footnotesize Figure SI1: Scattering intensities for the quadruplex solution at $C = 1.2$ mM, obtained from SAXS measurements and simulations (panels b – f). The accordance between experimental and simulated data has been evaluated through the residual sum of squares ($RSS$), calculated at different values of $T^*$ and $K$ (panel (a)).}}
  \label{fig:SI1}
\end{figure}

\begin{figure} [h!]
\centering
  \includegraphics[width=5.7in]{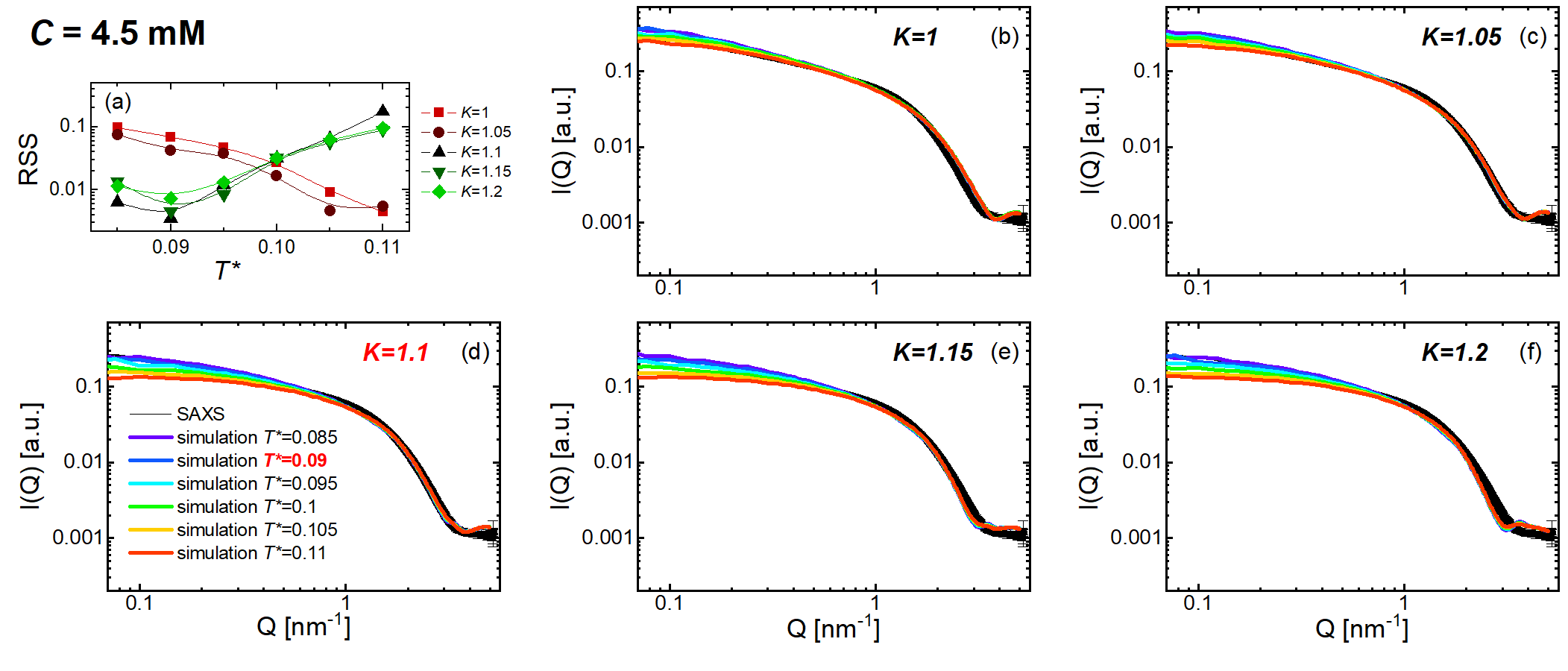}
  {\caption*{\footnotesize Figure SI2: Scattering intensities for the quadruplex solution at $C = 4.5$ mM, obtained from SAXS measurements and simulations (panels b – f). The accordance between experimental and simulated data has been evaluated through the residual sum of squares ($RSS$), calculated at different values of $T^*$ and $K$ (panel (a)).}}
  \label{fig:SI2}
\end{figure}

In Figure SI1 and SI2, the comparison between experimental and simulated scattering intensities for samples with DNA concentration respectively of 1.2 mM and 4.5 mM is reported. For $C=1.2$ mM, the best accordance is found when the simulation parameters assume the values of $K=$1.05 and $T^*=$0.1. For $C=4.5$ mM, the corresponding best parameters are $K=$1.1 and $T^*=$0.09.

\subsection*{2. Comparison between different estimates of the average dimension of G4 monomers and multimers}

\begin{figure} [h!]
    \centering
  \includegraphics[width=5in]{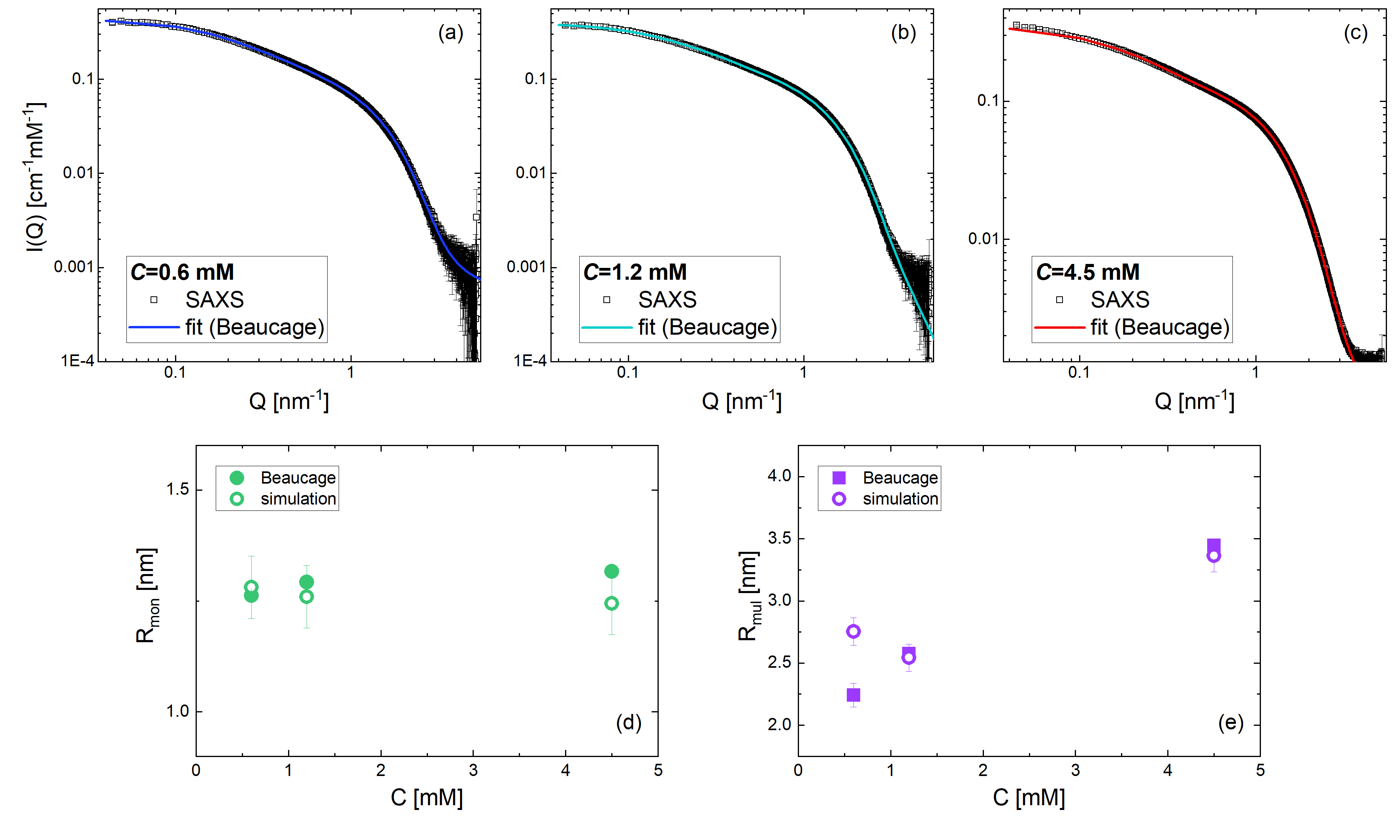}
  {\caption*{\footnotesize Figure SI3: Panel (a), (b) and (c) report the comparison between SAXS data and the corresponding fitting function according to Equation \ref{Beaucage}, at concentration $C=0.6$ mM, 1.2 mM and 4.5 mM respectively. Values of the radius of gyration of the G4 monomer ($R_{mon}$) and of the G4 multimers ($R_{mul}$) are compared to the corresponding values obtained by the simulations in panel (d) and (e) respectively.}}
  \label{fig:SI3}
\end{figure}

An estimate of the average size of G4 monomers and multimers can be obtained by fitting the experimental data using global scattering functions able to describe the behaviour of polymer chains and other structural systems with multiple structural levels \cite{Beaucage2004,Beaucage1996}:

\begin{equation}
\begin{split}
I(Q) \simeq & \ A_1 \text{exp}(-Q^2R^2_{agg}/3)+A_2\text{exp}(-Q^2R^2_{mul}/3)(1/Q^*)^{P_{mul}}\\
&+A_3\text{exp}(-Q^2R^2_{mon}/3)+A_4(1/Q^*_{mon})^{P_{mon}}+bck,
\label{Beaucage}
\end{split}
\end{equation}

where $Q^*=Q/[ \, \text{erf}(QkR_{agg}/6^{1/2}) ]^3 \,$ and $Q_{mon}^*=Q/[ \, \text{erf}(Qk_{mon}R_{mon}/6^{1/2}) ]^3 \,$, $A_i$ ($i=1,4$) are constant amplitudes and $bck$ gives the constant background. The first term in Equation \ref{Beaucage} describes the very low $Q$ region and takes into account large-scale scattering objects (aggregates of G4 multimers) with size $R_{agg}$. The second term describes the mass-scaling regime at intermediate $Q$ with two structural limits, which are given by the small-scale size $R_{mon}$ (related to the dimension of the monomers) and the intermediate-scale size $R_{mul}$ (related to the dimension of small multimers). The last two terms describe the behaviour of the quadruplex structural unit at high $Q$. The fitting parameters for the three concentrations are reported in Table SI1. The intensity curves are well described by a Porod behaviour for the G4 monomer ($P=4$), whereas the multimers approach the fractal behaviour of a polymer in good solvent conditions ($P_{mul} \approx 1.5$) \cite{Beaucage1996}.

 \begin{table}[h!]
 \centering
   \caption*{\footnotesize Table SI1: Parameters obtained by fitting the SAXS data by means of the expression reported in Equation 1.}
  \begin{tabular}{cccccccc}
    \hline
    $C$ [mM] & $R_{agg}$ [nm] & $R_{mul}$ [nm] & $R_{m}$ [nm] & $P^a$ & $P_{mon}$  & $k^b$ & $k_{mon}$\\
    \hline
    0.6 & 10.5 $\pm$ 0.2 & 2.24 $\pm$ 0.09 & 1.26 $\pm$ 0.01 & \multirow{3}{*}{1.33 $\pm$ 0.04} & \multirow{3}{*}{4} & \multirow{3}{*}{1.06} & \multirow{3}{*}{1.06} \\
    1.2 & 11.8 $\pm$ 0.03 & 2.58 $\pm$ 0.03 & 1.29 $\pm$ 0.01 & & & & \\
    4.5 & 13.6 $\pm$ 0.04 & 3.45 $\pm$ 0.02 & 1.32 $\pm$ 0.01 & & & & \\
    \hline
  \end{tabular}\\
\footnotesize{$^a$The value was obtained by fitting the low-concentration data and kept constant for the other two values of $C$. $^b$Values of $k$ and $k_{mon}$ where chosen according to Reference \cite{Beaucage1996}.}
 \label{tab:1}
\end{table}

In Figure SI3, the experimental data together with the corresponding fitting functions are reported. It is interesting to compare the values of $R_{mon}$ and $R_{mul}$ obtained by fitting the SAXS data with those evaluated by the simulations as $R_{mon}=L^2/12+R^2/2$ and $R_{mul}=(ML)^2/12+R^2/2$, where $L$ and $R \equiv D/2$ are the characteristic dimensions of the simulation HC and $M$ is the average multimer size. From Figure SI3(d) and (e), it can be observed how there is a good agreement between experiments and simulations. 

\subsection*{3. Estimation of the parameters describing the aggregation of single G4 conformations}
  
 \begin{table}[h!]
 \centering
  \begin{tabular}{cccccc}
    \hline
    G4 conformation & $L$ [nm] & $D$ [nm] & $T^*$ & $k_I$ & $\sigma_b$\\
    \hline
    Antiparallel & 5.3 & 1.4 & 0.21 & 2.752 & -4.88\\
    Hybrid & 3.4 & 2.2 & 0.05 & 3.038 & -5.80\\
    Parallel & 2.8 & 2.7 & 0.06 & 3.287 & -6.14\\
    \hline
  \end{tabular}\\
 \caption*{\footnotesize Table SI2: Parameters associated with each of the three possible quadruplex conformations.}
 \label{tab:1}
\end{table}

To obtain the parameters describing the stacking of single G4 conformations (antiparallel, hybrid, and parallel), we combined the information on the average size of the stacking cylinder and the average binding energy, obtained by simulations, to the information on the fraction $f$ of G4 in each conformation, obtained by circular dichroism. At each value of concentration $C$, we assumed that the values of $K$ and $T^*$ found by the simulations were given by the linear combination of the parameters associated with each of the three conformations weighted on the corresponding $f_i$,  i.e.:

\begin{equation}
    K=\sum_i f_i K_i
    \label{eq:K}
\end{equation} 

and 

\begin{equation}
    T^*=\sum_i f_i T^*_i
    \label{eq:T}
\end{equation}

where $i=$antiparallel, hybrid, and parallel (assuming that $K_i$ and $T^*_i$ do not depend on concentration). In Table SI2, the obtained values of $L$, $D$, and $T^*$ are reported.

\medskip

\printbibliography